\documentstyle[floats,twocolumn,prl,aps,psfig]{revtex}  

\begin{document} 

\twocolumn[{   
 
\draft 
\title{Spin and interaction effects on charge distribution and  
currents in one-dimensional conductors and rings within the 
Hartree-Fock approximation}   
%
\author{Avraham Cohen$^{1}$, Klaus Richter$^{2}$, and 
Richard Berkovits$^{1}$}
\address{$^1$The Minerva Center for the Physics of Mesoscopics, 
Fractals and Neural Networks, Department of Physics, \\
Bar-Illan University, 52900 Ramat-Gan, Israel \\ 
$^2$Max-Planck-Institut f\"ur Physik komplexer Systeme, 
N\"othnitzer Strasse 38, 01187 Dresden, Germany 
} 
\date{\today}
\maketitle 
\mediumtext
{\tighten
\begin{abstract}
Using the self--consistent Hartree-Fock approximation for 
electrons with spin at zero temperature, we study the effect of the electronic 
interactions on the charge distribution in a one-dimensional 
continuous ring containing a single $\delta $ scatterer. We reestablish 
that the interaction suppresses the decay of the Friedel oscillations. 
Based on this result, we show that in an infinite one dimensional conductor 
containing a weak scatterer, the current is totally suppressed because of 
a gap opened at the Fermi energy.  
In a canonical ensemble of continuous rings containing many scatterers,  
the interactions enhance the average and the typical persistent current.  
\end{abstract}
}

}]

\narrowtext 


\noindent 
The effects of electronic interactions on characteristic properties,  
such as charge fluctuations, persistent currents (PC's) and the conductance 
of electronic systems are very rich and interesting.\cite{reviews}    
They strongly depend on the strength and range 
of the interactions,\cite{apel82,gogolin94,egger95} on the  
dimensionality of the system, and on whether the space is discrete or 
continuous.\cite{shulz93,liebwu68}    
Approximate calculations, like Hartree-Fock,  
introduce a great deal of simplifications, but at the 
same time  many effects may be washed out.  
However,  approximate calculations may be used to shed more light 
on specific problems, while keeping in mind their limitations. 
In this work we consider $e$-$e$ interactions within 
the self-consistent Hartree-Fock approximation (SCHFA) 
for electrons 
with spin at zero temperature.  
For simplicity we assume an equal number of electrons of 
opposite spin states. 
%

Our aim is to  study numerically the interaction effects 
on the charge distribution and  the currents in continuous one-dimensional (1D)  
isolated rings and open conductors containing a single 
$\delta$ scatterer,\cite{egger95,kanefisher9292,glazman9394,fisherglazman96,pschmitt}  
as well as on the PC's in rings containing many scatterers.\cite{ba,by-l-ch-m}  
Even within the Hartree-Fock approximation we recover the 
bosonization\cite{egger95} and the density-matrix 
renormalization-group result:\cite{pschmitt}   
We show that for a single scatterer in a ring the 
repulsive electronic interaction suppresses the decay of 
the charge oscillations. 
Based on this we show, as a central result, that for an open 
conductor with a weak 
scatterer the electronic conduction at the Fermi energy vanishes because 
of Bragg reflection coexisting with a gap at the Fermi energy.
The zero conduction of the interacting system  was obtained 
in Refs.\onlinecite{kanefisher9292,glazman9394,fisherglazman96} 
by exact and by renormalization group calculations.  
Within the first iteration of the SCHFA, it was 
shown\cite{glazman9394,fisherglazman96} 
that an attempt to explain this result by a scattering 
perturbation series is inadequate  
because of logarithmic divergences of the transmission 
amplitude at the Fermi energy in all orders of the series. 

Although the dissipative conductance of the infinite conductor  
is suppressed by the interactions, the PC in a ring is not.  
This is because the conductance depends on the 
properties of the levels close to the Fermi energy 
but the PC is a thermodynamic 
property that depends on the response of all occupied levels.\cite{ba}  
Moreover, we show that once many scatterers are considered, the interactions 
not only do not suppress the PC, but even enhance it.  

\noindent 
We write the HF equation for electrons in a ring of radius $R$  
with angular coordinate $\theta$  and energy units 
$\hbar^2 / m_e R^2=1$ (we drop the background term) 
as   
\begin{eqnarray}
&  &-
\left[{1 \over 2} { \partial^2 \over \partial \theta^2} +
V_{\rm dis}(\theta) + {R\over r_0} 
\int_0^{2\pi}  { \sum_{l'=1}^{Ne} | \psi_{l'}(\theta') |^2
\over \sqrt{(\theta-\theta')^2+\epsilon^2}   } 
d{\theta'} \right] \psi_{l}(\theta)  
\nonumber \\
&  & -
\delta_{s_{l'},s_l} {R \over r_0}
\int_0^{2\pi}
{ \sum_{l'=1}^{Ne}\Psi_{l'}^{*}(\theta')\Psi_{l'}(\theta) 
\over \sqrt{(\theta-\theta')^2+\epsilon^2}   }
\psi_{l}(\theta')d\theta'
 =  
E \psi_{l}(\theta)    \, .
\label{hfintegro}
\end{eqnarray} 
The twisted boundary condition   
$\psi(\theta+2\pi)=\psi(\theta)\exp(i2\pi \phi / \phi_0)$ 
accounts for a flux $\phi$ threading the ring.   
$\phi_0\equiv hc/ e$ is the flux quantum.   
$V_{\rm dis}(\theta)$ is the disorder potential which may include a 
single or many scatterers.  
The first (second) integral term is the Hartree (Fock) term.  
The electronic wave functions 
$\Psi_l(\theta)\equiv\psi_l(\theta)\exp{( -i\theta \phi/\phi_0 )}$
in the Fock term are $2\pi $ periodic for any value of flux.  
$l$ enumerates the energy levels together with the spin state $s_l$. 
$N_e$ is the total number of electrons in the ring. 
The cutoff $\epsilon^2$  allows (as in quasi 1D) using the 3D 
Coulomb law\cite{shulz93} and
makes the integrations finite.   
The square of the distance between the particles is defined\cite{distance} by
$(\theta-\theta')^2 \equiv
\min [|\theta-\theta'|^2, (2\pi-|\theta-\theta'|)^2]$.  In Eq.~(1),
$r_0 \equiv {\bf \varepsilon} \hbar^2/m_e e^2$ denotes the Bohr radius
with dielectric constant ${\bf \varepsilon}$
(to be distinguished from the cutoff $\epsilon $). 
We define the coefficient $g \equiv {R/ r_0}$ to 
be the interaction strength. $g \sim 1$ corresponds to 
semiconductors.\cite{cohenbh}   
%
%
Because the sum $\sum_{l'=1}^{Ne}\Psi_{l'}^{*}(\theta')\Psi_{l'}(\theta)$ 
represents almost a closure relation we replace, 
as discussed in Refs.\onlinecite{cohenbh} and \onlinecite{uslandau}, 
the integrodifferential equation (Eq. (\ref{hfintegro})) by an  
ordinary Schr\"odinger equation 
that we solve {\em self--consistently}:
\begin{eqnarray}
\bigg[-{1 \over 2} { \partial^2 \over \partial \theta^2}
+ 
V_{\rm dis}(\theta) 
+
g V_{\rm eff}(\theta) \bigg] \psi_{l}(\theta)
= E \psi_{l}(\theta).  
\label{hfeq}
\end{eqnarray}  
Here  $V_{\rm eff}(\theta)$ is given by 
\begin{eqnarray}  
\int_0^{2\pi}  { \sum_{l'=1}^{Ne} | \psi_{l'}(\theta') |^2 -
\delta_{s_{l'},s_l} 
{\rm {\rm Re}} \{\Psi_{l'}^{*}(\theta')\Psi_{l'}(\theta)\}
\over \sqrt{(\theta-\theta')^2+\epsilon^2}   }
d\theta'
\nonumber 
\end{eqnarray} 
where {\rm Re} stands for real part. 
The spin degree of freedom is very important. 
For spinless electrons the interaction effect is weak because 
the Fock and Hartree terms tend to cancel each other due to 
opposite signs and similar absolute values. 
%
Taking into account the spin degree of freedom, the Hartree term is 
twice as large as the exchange term.
Then the former dominates  $V_{\rm eff}$ and enhances screening; 
therefore we expect the interaction effects to be stronger 
for electrons with spin.  
This explains the importance of considering spin\cite{gsh95assa95} in 
order to understand disordered interacting systems.
%

\begin{figure}
[t]
\psfig{figure=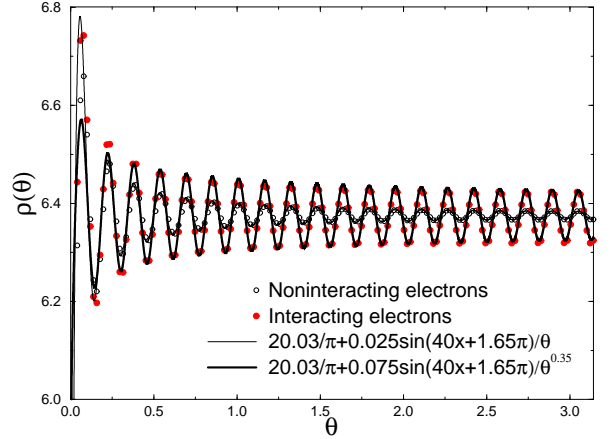,angle=-90,height=7.cm} 
\caption{
Charge oscillations per spin, along half circumference of a 1D 
continuous ring, 
induced by a {\em weak} single
scatterer $\lambda \delta (\theta)$   
($\lambda=3.8$). 
Thin (bold) symbols stand for noninteracting (interacting) electrons. 
The interactions, at self-consistency of the Hartree-Fock calculations, 
tend to make 
$\rho(\theta)$ periodic and to minimize $\rho(0)$ further. 
The interaction strength is $g=3$, and the flux threading the 
ring is $\phi / \phi_0=0.05$ (see text).
The total number of electrons per spin is 40. 
The curves are the estimations by the indicated formulas. 
}
\label{fig1}
\end{figure}

We begin by studying the interaction effect on the charge oscillations 
in a ring with a single scatterer,    
\begin{equation} 
V_{\rm dis}(\theta)= \lambda \delta(\theta).  
\label{vdis1}
\end{equation} 
For a strong scatterer, $\lambda \ge E_f$ ($E_f$ is the Fermi energy),  
the interaction effect on the decay of the charge oscillations  
is weak and may even be neglected because the scatterer is 
dominating. 
For a weak scatterer,  $\lambda<<E_f$,  
at the level of the SCHFA we recover the 
numerical result of Ref.\onlinecite{pschmitt} based on the density-matrix
renormalization group:
With increasing repulsive interaction $g$ the decay of the Friedel oscillations 
is suppressed (indicating also the reliability of our SCHFA).  
{Figure \ref{fig1}} depicts the decay rate for the 
strongest interaction for which the SCHFA still converges.  
As Fig. 2 shows, the effective potential tends to be periodic 
with half a Fermi wavelength periodicity.  
Both (direct and exchange) terms tend to have this periodicity 
which is independent of the interaction strength. 
This behavior holds for a larger number of
electrons on a ring for a given constant charge density. 
\begin{figure}
[t]
\psfig{figure=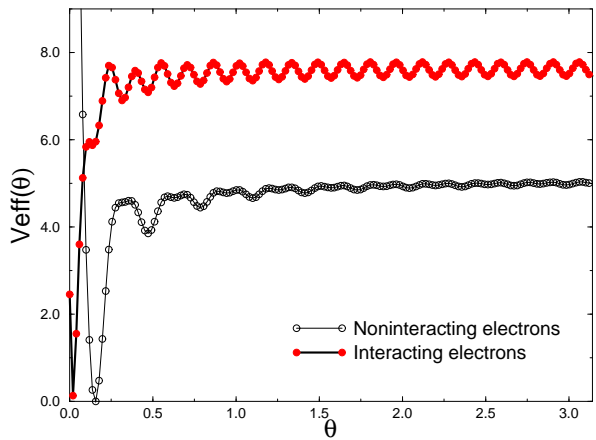,angle=-90,height=7.cm} 
\caption{ The effective potential   
 along the half circumference of the ring of Fig. 1. 
 The thin (bold) symbol is the noninteracting (interacting) result.  
 Note the clear tendency of $V_{\rm eff}(\theta)$ at self-consistency 
 to become periodic, and to screen the 
 scattering potential $\lambda \delta(\theta)$ (not shown). 
}
\label{fig2}
\end{figure} 
\begin{figure}[t]
\psfig{figure=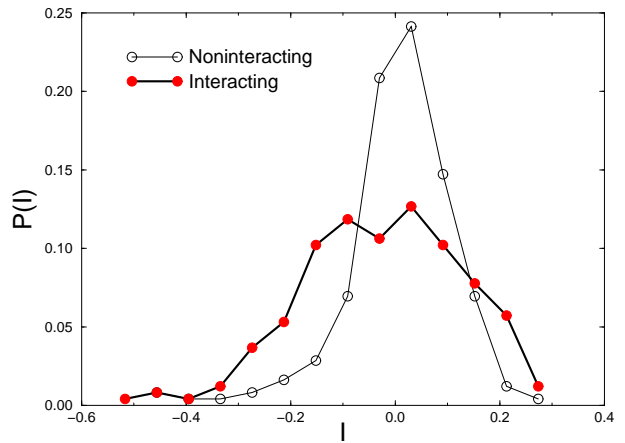,angle=-90,height=7.0cm} 
\caption{The interaction 
  effect on the statistics of the sample persistent 
  current (in units of the PC of the clean ring of noninteracting electrons). 
Thin (bold) symbols stand for noninteracting (interacting) electrons  for 
the same canonical ensemble of 201 realizations.  
The interaction ($g=1,\; \phi/\phi_0=0.325$) enhances the persistent current.
}
\label{fig3}
\end{figure} 
\begin{figure}[t]
\psfig{figure=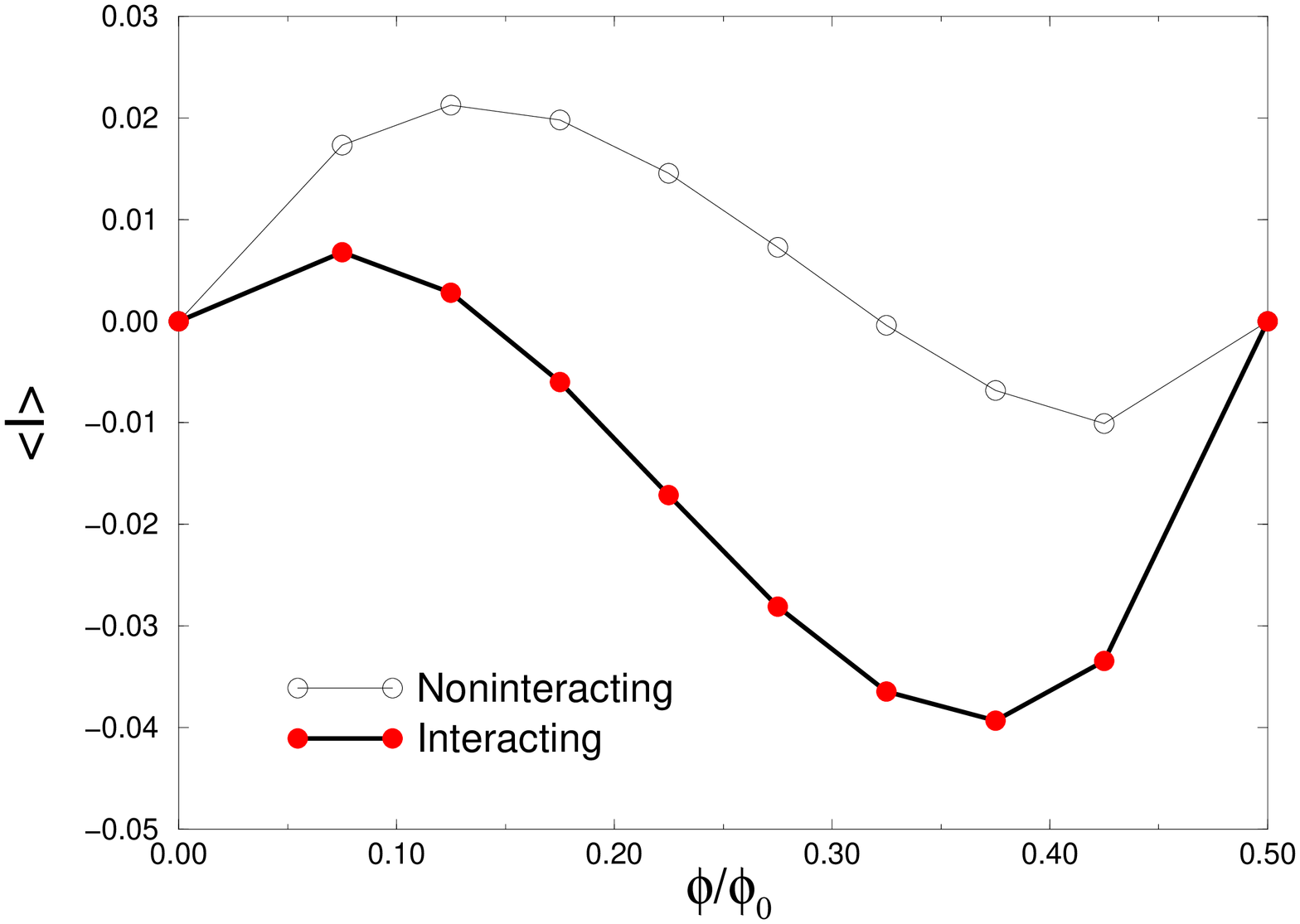,angle=-90,height=7.cm} 
\caption{
  The interaction ($g=1.75$)  
  enhances the average sample persistent current 
  (in units of the PC of the  clean ring) and introduces a 
  preffered diamagnetic current direction. 
  Thin (bold) symbols represent noninteracting (interacting) electrons 
  for the same 150 realizations. 
}
\label{fig4}
\end{figure} 
%

The above results may be used to study the effect on the charge 
oscillations and on the conduction in the case of a single weak scatterer 
$\hat{\lambda} \delta (x)$ embedded in an {\em infinite} 1D conductor 
($x$ is the spatial coordinate).   

For noninteracting electrons the orthogonal wave functions,   
with a given spin state, are\cite{glazman9394,fisherglazman96}    
\begin{equation}
\phi_k^{(1)} (x) =  
\left\{ 
\begin{array}{lr}
e^{i k x } + r(k,\lambda) e^{-ikx}, \; & \;  x < 0  \\ 
t(k,\lambda) e^{i k x},             \; & \;  x>0,     
\end{array} 
\right.
\label{phi1}
\end{equation} 
\begin{equation}
\phi_k^{(2)} (x) =  
\left\{ 
\begin{array}{lr}
t'(k,\lambda) e^{-i k x},             \; & \;  x < 0 \\ 
e^{-i k x } + r'(k,\lambda) e^{ikx},  \; & \;    x > 0,  
\end{array} 
\right.
\label{phi2}
\end{equation} 
with  $k>0$ and $\lambda \equiv \hat{\lambda} / (\hbar^2/m_e)$  
having units of inverse length.   
$ r(k,\lambda)={-i\lambda/ (k+i\lambda)} $  
and 
$  t(k,\lambda)={k /  (k+i\lambda)}.$ 
Because of time-reversal symmetry and the symmetry of  
the potential under coordinate inversion,
$ t'=t $ and $r'\equiv -(r/ t)^*t=r$.  
The fluctuating density per spin is 
\begin{eqnarray} 
\Delta \rho(x) &=&  
 2 \int_0^{K_f}{-\lambda^2\cos 2kx + \lambda k \; \sin 2k|x| \over
  k^2+\lambda^2}dk \nonumber \\ 
& =&  - 2 \lambda e^{2\lambda |x|} {\rm Im} \{E_1(-iz)\}.    
\label{rhor}
\end{eqnarray} 
${\rm Im}$ takes the imaginary part of the exponential integral $E_1$, 
$z\equiv 2|x|(K_f + i\lambda)$, and $K_f$ is the Fermi wave
vector.  
$\Delta \rho(0) = - 2 \lambda \tan^{-1} (K_f / \lambda)$    
is a minimum. For $K_f x > 1$, the asymptotic 
expansion of $ E_1$  
%
%
implies
\begin{equation} 
\Delta \rho(x) =  
 - {\lambda ( K_f \cos 2K_f x + \lambda \sin 2K_f |x| ) 
\over  |x|  (\lambda^2 + K_f^2) }.  
\label{rhor2}
\end{equation}  
Using $r\equiv |r|e^{i\eta}$ and
$|r_f|={|\lambda|  / \sqrt{K_{f}^2+\lambda^2}}$,   
$\sin\eta_f = {-k / \sqrt{K_{f}^2+\lambda^2}}$, 
one finds\cite{glazman9394,fisherglazman96}
\begin{equation} 
\Delta \rho(x)=  
{|r_f| \sin (2 K_f |x| + \eta_f )  \over |x|}.   
\label{drhopolarkuku} 
\end{equation} 
For the SCHFA the initial $V_{\rm eff}$ is calculated using the 
wave functions of noninteracting electrons.  
The charge fluctuations define the Hartree potential     
\begin{equation}  
 V_{_{H}}(x)= g_s
 \int_{0}^{\infty} \bigg[ {1 \over |x+x'|} + {1 \over |x-x'|} \bigg] 
\Delta \rho (x') dx',  
 \label{hartree1} 
\end{equation}
where $g_s=1 \; (2)$ for electrons without (with) spin. 
Our approximated Fock potential is   
\begin{eqnarray} 
V_{_{F}}(x) & = & - \int_{-\infty}^{+\infty} 
{   
\int_0^{K_f} 
\sum_{i=1}^{2}	 {\rm Re} 
\{ \phi_k^{(i)*}(x')  \phi_k^{(i)} (x)  
\}  dk  
\over |x-x'| 
}  dx' \nonumber \\ 
& = &      
 -
 \int_0^{\infty}  
\bigg[{1 \over |x+x'| } + {1 \over |x-x'| }\bigg] \Delta \rho (x+x')
dx' .     
\label{vexp7}  
\end{eqnarray} 
Clearly,  $ V_{\rm eff} = V_{_{H}} + V_{_{F}} $ is a function of $|x|$,  
and will change during the iterations until self-consistency 
is reached. 
$ V_{\rm eff} $ is small due to a weak coupling constant ($g \sim 1$).
%
At this point we invoke an approximate self-consistency by adopting a 
suppression\cite{egger95,pschmitt} of the decay of the Friedel oscillations,  
%
%
as was demonstrated above to be valid in the SCHFA.  
We substitute by hand the 
limit\cite{egger95} $\delta=0$ in
\begin{equation}
\Delta \rho (x) =   
{ |r_f| \sin (2K_f|x| + \eta_f )  \over |x|^{\delta} } 
\label{reckern} 
\end{equation}  
for Eqs. (\ref{hartree1}) and (\ref{vexp7}), assuming that this yields
a $V_{\rm eff}$  close to that from the SCHFA. 
To carry out the integration [in Eqs. (\ref{hartree1}) and (\ref{vexp7})],  
we use a   
{\em cutoff} that allows contributions only from    
$|x-x'| \ge \epsilon.$ 
This cutoff is equivalent to that used in Eq. (\ref{hfintegro}). 
%
%
%
For $K_f x \gg 1$, we then obtain, up to an additive constant,    
\begin{eqnarray} 
V_{eff}(x) = U 
[ g_s \sin (2K_fx+\eta_f) - \sin (4K_fx+\eta_f) ],      
\label{veffint7} 
\end{eqnarray} 
where $U\equiv - 2  |r_f|  {\bf c}_i(2K_f\epsilon)$, and 
${\bf c}_{i}$ is the cosine integral.   
Equation (\ref{veffint7}) shows that $V_{\rm eff}$  has two periodicities: 
${\lambda_f / 2}$ from the direct potential 
($\lambda_f \equiv {2\pi/ K_f}$),  
and ${\lambda_f / 4}$ from the exchange potential.  
The overall periodicity is given by the larger period.  
The electrons at the Fermi energy  exactly 
obey the Bragg condition\cite{kittel} for total reflection, i.e., 
\begin{equation}  
2 {\lambda_f \over 2} \sin {\pi\over 2} =n \lambda_f  \, .
\label{bragg}
\end{equation}  
All the states with $ |k|<K_f $ remain practically unaffected
by the weak and periodic  $V_{\rm eff}$.  
Note that consistently with Eq. (\ref{bragg})   
there is a gap\cite{kittel} of order $U$ at the Fermi energy. 
Thus the current vanishes at the Fermi energy. 

For a ring with a weak scatterer 
the interaction will  not destroy the PC 
even if the current at the Fermi energy (assuming a large ring) 
is totally suppressed by the periodic effective potential. 
This follows from the fact that all occupied levels contribute to the PC, 
except at $E_f$, where Eq. (\ref{bragg}) is assumed to be relevant.
 
In the following we will consider the general case of a large number
of scatterers in a ring. Figure 2 already shows the importance of screening for
a single scatterer. This indicates that screening is of 
particular relevance for the case of many random scatterers:   
\begin{equation} 
V_{\rm dis}(\theta)=\sum_{j=1}^{N_s} \lambda_j \delta(\theta-\theta_j). 
\label{vdisNs}
\end{equation}
Here the location and strength of the $j$th scatterer are uniformly 
distributed in $(0,2\pi)$ and  $(-\Lambda, \Lambda)$, respectively. 
%
%
$N_s$ is the total number of scatterers in a ring. 
For the numerics we use $\Lambda=14$ (in scaled units). 

The characteristic features of disordered noninteracting samples 
were, e.g., discussed by Imry and Shiren.\cite{shiren}     
For noninteracting electrons, the localization length\cite{cohenbh}  
at $E_f=200$ is $\xi\sim{\pi /  2}.$ 
This should reduce the average current in open conductors by a  
factor $\sim 1/50$. The average sample PC of noninteracting   
electrons was reduced by factor $\sim 1/40$ which is slightly greater  
than predicted for open conductors.  
The typical sample PC, $\sqrt{\langle I^2\rangle }$, 
was reduced by factor $\sim 1/10$ 
which indicates the importance of a statistical study. 
The fixed 
total number of electrons in a ring was $32\pm4$.

{Figure 3} shows the interaction effect on the sample PC 
statistics for an interaction coupling constant $g=1$. 
The interaction reduces the
peak, centered at zero, while broadening the distribution. 
Furthermore, the distribution gains more weight at negative 
values of the PC, which indicates a diamagnetic tendency. 
We found that the interaction enhances the 
typical PC (by factor $\sim$2); the average PC is
neither  enhanced nor suppressed. 
{Figure 4} shows  that for increasing interaction, $g=1.75$, the 
average PC is also enhanced by factor $\sim$2.   
Figures 3 and 4 both show a clear tendency of the interaction 
to enhance the PC for electrons with spin.   
For spinless electrons  
the PC was found\cite{cohenbh} to be rather unaffected by interaction.     
This shows an essential difference 
between models of electrons with or without spin.  In addition, 
a clear difference between tight-binding models and continuous 
models\cite{weiden} becomes apparent:  
In the former it was concluded,\cite{bzy}    
using exact diagonalization and the SCHFA, that switching on 
the $e$-$e$ interaction  in the regime of moderate disorder further 
suppresses the PC because of the Mott transition.\cite{mott82}  
%
%
In continuous models this transition appears to be irrelevant, 
since the continuous models correspond to tight-binding models at very 
low fillings.\cite{gsh95assa95,kanefisher9292} 
%

In conclusion, using the SCHFA in one-dimension, we showed the 
tendency of the electronic interaction to build up nondecaying charge 
oscillations in a ring containing a single weak scatterer. 
Adopting this result for an infinite conductor implies a periodic 
effective potential. 
The electronic conduction was shown to vanish, because of Bragg  
reflection that coexists with a gap at the Fermi energy. 
This shows that, even in the HF limit the influence of the 
interactions on the Friedel oscillations and on conduction in 
one-dimension, calculated by exact and renormalization methods, 
may be reproduced. 
In rings the PC is not suppressed by the interaction. 
It is even enhanced in the case of many 
moderate scatterers due to screening. 
To demonstrate these effects,   
we considered the spin degree of freedom and used continuous 
conductors and rings. 

%

A. C. would like to thank A. Auerbach, D. Bar-Moshe, and  B. Shapiro for 
valuable discussions, and A. Heinrich for his interest in this work. 
A. C. and K. R. would like to thank U. Eckern, P. Schwab and 
P. Schmitteckert for valuable comments and criticism. 



%
%
%
%
%
%


\begin{references}
\bibitem{distance} This definition of the distance is close to 
$2 \; \sin({|\theta-\theta'|/ 2}) $  and  
does not introduce any qualitative changes. 
Also, the small cutoff has a negligible influence on the
numerical results.
\end{references}
\end{document}